\documentclass[a4paper,14pt]{article} 
\usepackage{graphicx} 
\usepackage{graphics} 
\usepackage{amsmath} 
\usepackage{amsfonts}  
\usepackage{amssymb} 

\usepackage[colorlinks=true, pdfstartview=FitV, linkcolor=blue,  citecolor=blue, urlcolor=blue]{hyperref}  

\newcommand{\Cs}{\mathbb{C}} 
\newcommand{\Us}{\mathbb{U}} 
\newcommand{\Vs}{\mathbb{V}} 
\newcommand{\Ws}{\mathbb{W}} 
\newcommand{\Pl}{\mathbb{P}} 

\def\l{\left}
\def\r{\right}
\def\beq{\begin{equation}}
\def\eeq{\end{equation}}

\begin{document}

\title{Scale relativity of the proton radius: solving the puzzle}

\author{Laurent Nottale\\
{{\small \it LUTH, Observatoire de Paris-Meudon, F-92195 Meudon, France}}}
\maketitle

\begin{abstract}
The proton size has been found, with a $6\,\sigma$ statistical significance, to be larger by 4\% when it is measured relatively to the electron than to the muon \cite{Pohl2010,Antognini2013}. We solve this proton radius puzzle by accounting for the relativity of the proton scale. The proton to electron and proton to muon scale ratios are obtained by direct measurement, but their comparison requires a conversion to electron reference which is currently made by assuming the usual law of scale ratio composition, $\rho_ {pe}=\rho_ {p\mu} \times \rho_ {\mu e}$. Using instead the special scale relativistic law $\ln \rho_ {pe}=(\ln\rho_ {p\mu}+\ln\rho_ {\mu e})/(\ln\rho_ {p\mu} \ln\rho_ {\mu e}/(\ln\rho_ {\mathbb{P} e})^2$, where $\mathbb{P}$ denotes the Planck length-scale, the two determinations of the proton radius, showing now a ratio $1.009 \pm 0.008$, recover their agreement within about $1\sigma$. The proton radius puzzle therefore provides one with a highly significant test of the special scale relativity theory.
\end{abstract}

 The ``proton radius puzzle" is a statistically highly significant (to 6 $\sigma$) discrepancy between the measurements of the proton charge radius using as reference scales the electron or the muon \cite{Pohl2010}. Such a problem directly involves the relativity of scales, since the two references for the proton size measurements differ by a factor $m_\mu/m_e \approx 207$.
 As we shall see, it can therefore be solved by accounting for special scale-relativistic corrections in the calculation of scale ratios \cite{Nottale1992,Nottale1993,Nottale2011}. 
 
 Measurements of the proton radius from laser spectroscopy of muonic hydrogen have shown up to six standard deviations smaller values than obtained from electron-proton scattering and hydrogen spectroscopy \cite{Krauth2017}. The method consists of using the proton size contribution to the Lamb shift in order to measure the extension of the charge distribution in the proton. 
 
 Recall that the main contribution to the Lamb shift comes from the charge radius of the electron. According to QED an electron continuously emits and absorbs virtual photons and as a result its electric charge is spread over a finite volume instead of being pointlike \cite{Eides2001}. 
 
 The resulting energy shift (see e.g. \cite{Eides2001} and references therein) may be written as
$E_L=\frac{1}{6} R_y ({r_e}/{r_{B}})^2$,
where $R_y=\frac{1}{2} \alpha^2 m_e c^2$ is the Rydberg constant and $r_{B}=\alpha^{-1}  \lambda_e$ is the Bohr radius of the electronic hydrogen.
 The proton size contribution to the Lamb shift is fundamentally of the same nature. Namely, it comes from the finite size $r_p=<r^2>^{1/2}$ of the charge distribution in the proton. Its contribution to the Lamb shift is therefore
$E_p=\frac{1}{6} R_y ({r_p}/{r_{B}})^2$, 
 in which $m_e$ is replaced by the reduced mass of the electron in the hydrogen atom. This effect is used to measure the proton radius from spectroscopy \cite{Mohr2016}.

 Muonic hydrogen has two main special features as compared with the ordinary electronic hydrogen atom. First, the role of the radiative corrections generated by the closed electron loops is greatly enhanced, and second, the leading proton size contribution becomes the second largest individual contribution to the energy shifts after the polarization correction \cite{Eides2001}. Its resulting Lamb energy shift between the $2S_{1/2}$ and $2P_{1/2}$ states has been theoretically found to be \cite{Pohl2010,Antognini2013}
 \beq
 E_{L\mu }=[206.0336(15) -5.2275(10) \: r_{p\mu}^2+0.0332(20)] \;{\rm meV},
 \eeq
 where $r_{p\mu}$ is expressed in fm ($10^{-15}$m) and where the last term is the two-photon exchange contribution. From the measurement of the Lamb shift in muonic hydrogen, a value \cite{Pohl2010,Antognini2013}
 \beq
 r_{p\mu}=0.84087(39) \;{\rm fm}
 \eeq
 has been found, while the 2014 CODATA-recommended value based on electronic hydrogen is \cite{Mohr2016}
 \beq
 r_{pe}=0.8751(61) \;{\rm fm}.
 \eeq
Therefore, there is a more than $5\sigma$ discrepancy between the relative-to-electron and relative-to-muon determinations. The ratio of these two determinations is $\rho_{e \mu}=1.041 \pm 0.008$.

The proton radius discrepancy has been more recently confirmed and reinforced by a measurement of the deuteron radius in muonic deuterium \cite{Pohl2016}. The value of the proton charge radius derived is once again low, $r_{dp}=0.8356(20)$ fm, in statistical agreement with the muonic hydrogen result. On the other side, the CODATA  value of the electron-based proton radius has been recently supported by a new measurement of the 1S-3S transition frequency of hydrogen, yielding $r_{pe}=0.877(13)$ fm \cite{Fleurbaey2018}, although a smaller value has also been found from a measurement of the 2S-4P transition \cite{Beyer2017}.  These new measurements need to be integrated in a global adjustement before being used in a comparison with theoretical predictions.

 We shall now show that this discrepancy disappears in the framework of the special scale relativity theory. Recall that the principle of scale relativity \cite{Nottale1992, Nottale1993, Nottale2011, Nottale2019} (which is added to the current principle of relativity involving position, orientation and motion) states that there is no absolute scale in nature and that only scale {\em ratios} do have physical meaning.  
 
 The scale relativity theory is the general framework built from this first principle \cite{Nottale1993,Nottale2011}, including the construction of new scale laws of log-Lorentzian form \cite{Nottale1992}, a geometric foundation of quantum mechanics \cite{Nottale2007} and of gauge fields \cite{Nottale2006} based on a nondifferentiable and continuous (therefore fractal) space-time, and the suggestion of the possibility of a new macroscopic quantum-type mechanics based on a constant different from $\hbar$ and relevant to chaotic systems beyond their horizon of predictability \cite{Nottale1993,Turner2015,Nottale2018}.
 
 Here we are concerned with only the new scale laws aspect of the theory. Applied to the proton radius puzzle, the principle of scale relativity implies that the proton scale is not absolute, but depends on the reference scale which is used to measure it, respectively the electron scale and muon scale.
 
 We have mathematically proved \cite{Nottale1992}, \cite[Chapt.~6]{Nottale1993}, \cite[Chapt.~4.4]{Nottale2011} that the general solution to the special relativity problem (i.e., find the linear laws of transformations which come under this principle) is, as well for motion as for scales, the Lorentz transformation. This proof is based on only two axioms, internality of the composition law and reflexion invariance, which are both expressions of the only principle of relativity. We know since Poincar\'e and Einstein that the special motion relativity law of composition of two velocities $u$ and $v$ writes
 $w={(u+v)}/{(1+u \, v / c^2)}$.
 In the same way, the general law of composition of length-scale ratios writes in special scale relativity theory:
 \beq
 \Ws=\frac{\Us+\Vs}{1+\Us \Vs/\Cs^2},
 \eeq
where $\Us=\ln(r/\lambda)$, $\Vs=\ln \rho$ and $\Ws=\ln(r'/\lambda)$. This law yields the result $r'$ when applying a factor $\rho$ to a scale $r$ using a reference scale $\lambda$. The usual law $r'=\rho \times r$, i.e. $\ln (r'/\lambda)=\ln \rho + \ln (r/\lambda)$, which is actually  independent of any reference scale, is recovered in the limit $\Cs \to \infty$. The meaning of this constant can be clarified by expressing it also in terms of the reference scale $\lambda$:
\beq
\Cs_\lambda=\ln \frac{\lambda}{\lambda_\Pl}.
\eeq
 This introduces a new scale $\lambda_\Pl$ which is invariant under dilations and contractions, unreachable and unpassable, whatever the scale $\lambda$ which has been taken as reference. This remarkable property has naturally led us \cite{Nottale1992} to identify it with the Planck length-scale,
 \beq
 \lambda_\Pl=\sqrt{  \frac{\hbar \: G}{c^3}  }.
 \eeq
 However, it is clear that, at our scales, the standard laws of scale dilation $r'=\rho \times r$ (which corresponds to a Galileo group of transformation once expressed in logarithm form) are valid. Assuming that the new law (corresponding to a Lorentz group of scale transformations) is valid at small scales toward the Planck scale, there must exist a relative transition between Galilean scale relativity (GSR) and Lorentzian scale relativity (LSR). A natural identification of this transition is with the Compton scale of elementary particles, then, in the first place, of the electron \cite{Nottale1992}. Indeed, physics changes drastically at scales smaller than $\lambda_e$, in a way that is directly related to our purpose: namely, the various physical quantities, in particular masses and charges, become explicitly dependent on scale, a behaviour that is currently explained in terms of vacuum polarization and radiative corrections and well described by the renormalization group equations.
 
 With this identification of the reference scale, one obtains a numerical value of the constant $\Cs$ (which plays for scales a role similar to that of the velocity of light in vacuum for motion):
 \beq
 \Cs_e= \ln \frac{\lambda_e}{\lambda_\Pl}= 51.528.
 \eeq
 
 Let us now apply this framework to the proton radius puzzle. For this purpose, we must first analyse the way the proton radius calculations have been done so far. The final scale to which the result is refered is our macroscopic unit, $\lambda_u=1$ m. Since the scale relativity laws remain Galilean down to the Compton scale of the electron $\lambda_e$ , i.e., $r_p/\lambda_u=(r_p/\lambda_e) \times (\lambda_e/\lambda_u)$, it can be taken as reference in an equivalent way. 
 But we assume here that the law of composition of scale ratios encounters a transition from Galilean to Lorentzian scale laws at the scale $\lambda_e$.
 
In the electronic hydrogen experiment, the proton radius is measured with a reference scale which can be brought back to the electron Compton length, i.e. in terms of $(r_{pe}/\lambda_e)$. In the muonic hydrogen experiment, it is measured with a reference scale which is now the Compton length of the muon (207 times smaller), i.e. in terms of $(r_{p\mu}/\lambda_\mu)$. This is made clear by looking at the proton size contribution to the Lamb shift \cite{Eides2001} from which the proton radius is deduced, which may be written under the form:
\beq
E_p=\frac{1}{12} \alpha^4 m c^2 \l(\frac{r_p}{\lambda_c}\r)^2,
\label{Eq8}
\eeq
(where one uses the reduced mass in $m$ and $\lambda_c$). This expression holds as well for the electron as for the muon and shows in an explicit way that their Compton lengths $\lambda_c= \lambda_e$ or $\lambda_\mu$ are the natural reference length-scales in these measurements.

Now, in order to compare the two results, one needs to refer the proton radius to the same reference, i.e., $\lambda_e$. This is currently made, not by a measurement, but by a calculation which assumes implicitly the validity of Galilean scale laws, $r_{p\mu}/\lambda_e= (r_{p\mu}/\lambda_\mu) \times (\lambda_\mu/\lambda_e)$. 

However, beyond the electron Compton scale, one should use the special scale relativity law of composition. Before doing that explicitly, a last question should be solved: to which precise scale characterizing the proton charge extension should we apply them ? The energy shifts have been expressed in terms of the r.m.s. of distance calculated from the 3D charge density distribution in the proton,
\beq
r_p^2=<r^2>= \int_0^\infty 4 \pi r^2 \rho(r) \times r^2 dr.
\eeq
The proton charge density is known for long to be close to exponential \cite{Hofstadter1958},
$\rho(r)= \rho_0 \: e^{-r/a_p}$. The relation between the 3D ``radius" and the linear scale $a_p$ is:
 \beq
 a_p^2= \frac{1}{12}r_p^2.
 \eeq
The proton size contribution to the Lamb shift (Eq.~\ref{Eq8}) then takes an even simpler form:
\beq
E_p= \alpha^4  \: m c^2 \l(\frac{a_p}{\lambda_c}\r)^2.
\eeq 
This scale $a_p$, through its definition in a Yukawa-like behaviour and its 1D nature, is better adapted to characterize the proton size in connection with the electron and muon Compton lengths. This is supported by the fact that it is very close to the proton Compton length, $\lambda_p=\hbar/m_p c=0.83 \: a_p= 0.21031$ fm.

Finally, we can calculate the ``true" length-scale of the proton $a_{p \mu}^L$ referenced to large scales (equal or larger than $\lambda_e$) deduced from muonic hydrogen in the special scale relativity framework (involving log-Lorentzian scale laws). It now differs from the assumed value $a_{p \mu}^G=r_{p \mu}/\sqrt{12}$ calculated in the current framework of log-Galilean laws.  We set $\Ws_{ep}=\ln(\lambda_e/a_{p \mu}^L)$, $\Us_{e\mu} = \ln(\lambda_e/\lambda_\mu)=\ln(m_\mu/m_e)$, $\Vs_{\mu p}= \ln(\lambda_\mu/a_{p \mu}^G)$ and $\Vs_{ep}=\Us_{e\mu}+\Vs_{\mu p}=\ln(\lambda_e/a_{p\mu}^G) $. The correct ratio between the proton and electron scales derived from the muonium experiment is given by:
\beq
\Ws_{ep} = \frac{\Vs_{ep}}{1+\Us_{e\mu} \Vs_{\mu p}/\Cs_e^2}.
\eeq
We already see on this formula  that the expected scale ratio is smaller than its scale-Galilean counterpart, and therefore that the proton radius value refered to the electron will be increased.
Remark also that, since $\Us$ and $\Vs \ll \Cs$, the final difference between the log-Galilean and log-Lorentzian result (which is to be compared to the observed $4\%$ discrepancy) can be approximated by
$\delta \varrho = \Us_{e\mu} \Vs_{\mu p} \Vs_{ep} / {\Cs_e^2} $.  

With the numerical values $\Us_{e\mu}=5.332$, $\Vs_{\mu p}=2.040$ and $\Cs_e=51.528$, one finds
\beq
\varrho=1.031 \;\;({\rm from} \; a_p), \;\;\;  1.033 \;\;({\rm from} \; \lambda_p),
\eeq
which lies at $\approx  1 \, \sigma$ of the experimental ratio $1.041 \pm 0.008$. In other words, the proton size values refered to our scales now agree whether it is measured relatively to the electron [$r_{pe}= 0.8751(61)$ fm] or to the muon [$r_{p \mu}^L=(1.0320 \pm 0.0014) \: r_{p \mu}=0.8678(12)(4)$ fm], once the scale-relativistic corrections are taken into account.  In this result, the first uncertainty comes from the choice of the proton size in the composition law and the second is the experimental measurement uncertainty. The muon-based result now agrees with the electron-based measurement, but remains 5 times mor precise.
Another uncertainty comes from the fact that length-scales and mass-scales are no longer strictly inverse in the SSR framework. Their relation involves a Lorentz factor \cite[Chapt.~11]{Nottale2011} so that, while $\ln(m_\mu/m_e)=5.332$ one finds $\ln(\lambda_e / \lambda_\mu)=5.303$, but this correction remains negligible here owing to the other uncertainties.

To conclude, we note that besides the solution it brings to the proton radius puzzle, this result, if confirmed, puts to the test the special scale relativity theory and supports it in a highly significant way. This would be all the more remarkable as the new composition law is based on the Planck length which lies at scales $\approx 10^{22}$ times smaller than the electron scale, and is nevertheless identified here as an unreachable and unpassable scale, invariant under dilations. This is rendered possible by the fact that the scale variables are naturally logarithmic and that, as a consequence, the muon/electron ratio $\approx 200$ yields  $\Vs/\Cs \approx 2.3/22$ (in decimal logarithm) $\approx 1/10$, an already large ``relativistic" factor, and by the extraordinary precision reached by both theoretical and experimental studies of the Lamb shift.

Many indirect evidences of this new status of the Planck spacetime scale had been pointed out \cite{Nottale1992,Nottale1993,Nottale2011}:  for example, masses and charges in quantum field theories become finite at infinite energy; the electric charge takes its natural value $1/2 \pi$ at this limit; the U(1), SU(2) and SU(3) coupling constants converge together and with the gravitational coupling at the Planck energy scale, suggesting a direct Grand Unification at this scale, including gravitation. But the proton radius puzzle has now provided us with a direct and highly significant quantitative test of the theory, since it scans the scale ratios themselves and their composition law between electron, muon and proton scales.




   
\end{document}